\documentclass{article}
\usepackage{amssymb}
\usepackage{amsmath}
\usepackage{amsfonts}
\usepackage{epsfig}
\usepackage{graphicx}

\newtheorem{theorem}{Theorem}

\newtheorem{proposition}[theorem]{Proposition}

\begin{document}

\title{Nonabelian lattice theories: Consistent measures and strata}
\author{\textit{R.Vilela Mendes}\thanks{%
rvilela.mendes@gmail.com; rvmendes@fc.ul.pt; http://label2.ist.utl.pt/vilela/%
} \\
CMAFCIO, Faculdade de Ci\^{e}ncias, Universidade de Lisboa}
\date{ }
\maketitle

\begin{abstract}
The role of consistent measures in the rigorous construction of nonabelian
lattice theories is analized. General conditions that measures must fulfill
to insure consistency, positivity and a mass gap are obtained. The impact of
nongeneric strata on the nature of the Hamiltonian lattice potential is also
discussed.
\end{abstract}

\section{Introduction}

There are two ways to look at lattice theories. In one of them, to
characterize a theory in the continuum one starts from a discretized
spacetime, with a finite lattice of points and then, by successively
decreasing the lattice spacing $a$ and increasing the total volume $V$ of
the lattice, one approaches the continuum theory. At each step, taking an
Euclidean point of view, the discrete approximation of theory rests defined
by the choice of a measure. There is no need for these intermediate measures
to have a well defined physical meaning. All that is required is that in the
limit $a\rightarrow 0$ and $V\rightarrow \infty $ one obtains the desired
continuum measure. Of course, because in the limit one is dealing with the
delicate problem of measures in infinite dimensional spacetimes, the choice
of the intermediate finite dimensional measures turns out to be important to
insure the existence of the limit measure. However this is only a
requirement of mathematical convenience.

An alternative way to look at lattice theories is to consider the lattice as
an observational scaffold of some unknown theory which may or may not be
defined in the continuum, either because it possesses some intrinsic
validity cutoff or even because the spacetime manifold changes its nature at
small distances \cite{VilelaJPA} \cite{VilelaJMPNC} \cite{VilelaPRD}. In
this latter case the definition of the lattice and its associate measure is
much less arbitrary, because at each step one requires an exact description
of the theory at the length scale of the lattice.

In both instances, whether as a device to approach the continuum limit or a
scaffold to define the theory at all length scales, the mathematical notion
of \textit{consistent measure }(to be defined in the next subsection) is the
appropriate tool. In the first case because it insures the existence of a
limit measure and in the second because it insures probabilistic consistency
in the description of the physical system at all length scales.

Another important point in the formulation of nonabelian gauge theories is
the fact that in these theories the configuration space as well as the phase
space are not manifolds but orbifolds with singular points corresponding to
states of higher symmetry \cite{Cobra} \cite{Emmrich}. Therefore the
characterization of their \textit{strata} is important in particular because
they might correspond to a multiplicity of condensate backgrounds or
potential minima. A multiplicity of backgrounds might lead to a multiplicity
of solutions in the (Schwinger-Dyson) gap equation for fermion mass
generation \cite{Triantaphyllou} \cite{Blumhofer}. The nature of the gauge
strata in the nonabelian lattice theories will be analyzed. Because quantum
wave functions naturally explore different strata, the impact of nongeneric
strata on the backgrounds spectrum is related to the nature of the lattice
potential associated to the measure.

\subsection{Projective limits and consistent measures as tools to approach
the continuum or to describe physical systems at all scales}

The definition of consistent measures on a sequence of spaces (lattices in
this case) requires that the sequence be defined in an appropriate way. This
leads to the notion of:

\textbf{Projective limit}

Let $(I,\preceq )$ be a directed\footnote{%
Partially ordered and for any $i_{1},i_{2}$ in $I$, there is an $i_{d}$ in $%
I $ such that $i_{1}\preceq i_{d}$ and $i_{2}\preceq i_{d}$} set and $%
\left\{ X_{i};i\in I\right\} $ a family of topological separable spaces
indexed by $I $. For every pair $i,j\in I$ with $i\preceq j$ there is a
continuous map $\pi _{ij}:X_{j}\rightarrow X_{i}$ such that%
\begin{equation}
\pi _{ij}\circ \pi _{jk}=\pi _{ik}\hspace{1cm}\text{for\hspace{1cm}}i\preceq
j\preceq k  \label{PL1}
\end{equation}%
This is called a \textbf{projective family of spaces}.

The \textit{projective limit} of the family is the subset $X$ of elements of
the Cartesian product $\Pi _{i\in I}X_{i}$ such that its elements $x=\left(
x_{i}\right) $ satisfy%
\begin{equation}
x_{i}=\pi _{ij}x_{j}\hspace{1cm}\text{for\hspace{1cm}}i\preceq j  \label{PL2}
\end{equation}%
Furthermore there are mappings $\pi _{i}:X\rightarrow X_{i}$ such that%
\begin{equation}
\pi _{i}=\pi _{ij}\circ \pi _{j}\hspace{1cm}\text{for\hspace{1cm}}i\preceq j
\label{PL3}
\end{equation}%
The projective limit $X$ is denoted%
\begin{equation}
X=\underleftarrow{\lim }X_{i}  \label{PL4}
\end{equation}

\textbf{Consistent measures}

Of special interest is the case where each space $X_{i}$ is equipped with a
probability measure $\mu _{i}$. The family of measures $\left\{ \mu
_{i}\right\} $ is called a \textit{consistent} family if for every
measurable set $A$ in $X_{i}$ 
\begin{equation}
\mu _{i}\left( A\right) =\mu _{j}\left( \pi _{ij}^{-1}\left( A\right)
\right) \hspace{1cm}\text{for\hspace{1cm}}i\preceq j  \label{PL5}
\end{equation}

\textbf{The measure on the projective limit}

The next question is to know when, given a consistent family of measures,
there is also a measure $\mu $ on the projective limit $X$. The most general
requirement is probably the following \cite{Kisynski} \cite{Maurin}:

\textit{For every }$\varepsilon >0$ \textit{there is a compact subset }$%
K\subset X$ \textit{such that}%
\begin{equation}
\mu _{i}\left( X_{i}\setminus \pi _{i}\left( K\right) \right) \leq
\varepsilon \hspace{1cm}\text{for every }i\in I  \label{PL6}
\end{equation}%
and $\mu $ is denoted%
\begin{equation}
\mu =\underleftarrow{\lim }\mu _{i}  \label{PL7}
\end{equation}%
The existence of an infinite-dimensional measure as a limit of consistent
finite-dimensional measures is a powerful concept in the sense that the
limit measure might even be of different nature from the finite-dimensional
ones, for example not being absolutely continuous in relation to the
reference measure of the finite-dimensional ones.

A classical illustration of these concepts is Kolmogorov's construction of a
stochastic process $\left\{ X_{t}:t\in \mathbb{R}^{+}\right\} $ from its
finite-dimensional distributions \cite{Rao}:

Here $I=\mathbb{R}^{+}$, the $X_{i}$ spaces are discrete subsets $\left\{
t_{1},t_{2},\cdots ,t_{n}\right\} $ of $\mathbb{R}^{+}$ ordered by inclusion
with measures $\mu _{t_{1},t_{2},\cdots ,t_{n}}$ being the
finite-dimensional probability distributions. The mappings $\pi _{ij}$ are
simply the coordinate projections from the sets $\left\{ t_{1},t_{2},\cdots
,t_{n}\right\} $ to its subsets. One important point to notice in this
construction is that the projective limit (a subset of the Cartesian
product) is not simply the continuum limit of the stochastic process,
instead it also contains, in a consistent manner, the description of the
process at all levels of observation.

In this paper this framework will be applied to the description of physical
systems extended in space-time, that is \textit{fields}. Therefore it is
natural to consider the directed set to be a space-time lattice that, by
subdivision of its length elements, is successively refined and ordered by
inclusion. Notice that there is no need to identify the lattice with the
physical system, the lattice might simply be a \textit{observational scaffold%
} of the physical system at successively smaller scales. There is also no
need to assume that the physical systems are defined up to arbitrarily
smaller scales. The directed set $I$ may very well stop at some nonzero
scale. The definition of the sequence of spaces and their mappings is not
particularly difficult. The important point is of course to find the
consistent measures relevant to each physical system.

\subsection{Heat kernels}

Heat kernels are wonderful mathematical objects and, as has been pointed out
before \cite{ubiq1} \cite{ubiq2}, appear in all manner of places and
disguises. But where does the wonderfulness of heat kernels come from?

Given the operator $-O_{x}+\frac{\partial }{\partial t}$ where the operator $%
O$ is an elliptic operator, in particular a Laplacian in $\mathbb{R}^{n}$,
on a manifold or on a group manifold, the heat kernel is the solution of the
equation%
\begin{equation}
\left( -O_{x}+\frac{\partial }{\partial t}\right) K_{t}\left( x,y\right) =0
\label{HK1}
\end{equation}%
What makes the heat kernel a powerful gadget is the convolution\footnote{%
Convolution defined as $\int K_{t}\left( x-y\right) f\left( y\right) d\mu
\left( y\right) $ or $\int K_{t}\left( xy^{-1}\right) f\left( y\right) d\mu
_{H}\left( y\right) $ for a group} approximating property%
\begin{equation}
\lim_{n\rightarrow \infty }K_{\frac{1}{n}}\ast f=f  \label{HK2}
\end{equation}%
$f$ being a continuous and bounded function, as well as its representation
as a theta series%
\begin{equation}
K\left( t,x,y\right) =\sum_{k}\phi _{k}\left( x\right) \phi _{k}\left(
y\right) e^{-\lambda _{k}t}  \label{HK3}
\end{equation}%
for a manifold, with the $\phi _{k}^{\prime }s$ being a complete orthonormal
set of eigenvectors of the Laplacian and $\left\{ \lambda _{k}\right\} $ the
set of their eigenvalues or a similar expression for a group manifold with $%
x $ and $y$ group elements and the $\phi _{k}^{\prime }s$ representation
characters. Another important property, which naturally follows from the
evolution equation (\ref{HK1}) is the convolution semigroup property%
\begin{equation}
K_{t_{1}}\ast K_{t_{2}}=K_{t_{1}+t_{2}}  \label{HK4}
\end{equation}%
This property turns out to be of critical importance for the construction of
consistent measures. Notice also that a large family of convolution
semigroups of the heat kernel type do exist not only associated to
Laplacians but also to other more general L\'{e}vy kernels \cite{Bogdan}.

\subsection{A projective lattice family for spacetime fields}

In $n$-dimensional spacetime consider a set $\left( \mathcal{L},\preceq
\right) $ of successively finer hypercubic lattices ordered by inclusion.
Starting from some initial hypercubic lattice $L_{0}$ with lattice spacing $%
a_{0}$, the successive elements $L_{i}\in \mathcal{L}$ in this ordered set
are obtained both by regular subdivision of already existent plaquettes as
well as by the addition of new square or rectangular plaquettes. For the
purpose of definition of the mappings $\pi _{ij}:L_{j}\rightarrow L_{i}$
with $i\preceq j$ the elementary sets are the plaquettes, the plaquettes in $%
L_{j}$ being mapped on the corresponding plaquettes on $L_{i}$ or on the
empty set when they are new plaquettes which do not correspond to a
subdivision of the plaquettes in $L_{i}$. (this is analogous to the
coordinate projections in the Kolmogorov construction). The property $\pi
_{ij}\circ \pi _{jk}=\pi _{ik}$ for $i\preceq j\preceq k$ being verified, a
projective family of lattices is obtained, with projective limit defined as
the subset $L$ of elements of the Cartesian product $\Pi _{i\in I}L_{i}$
such that its elements $L=\left( L_{i}\right) $ satisfy%
\begin{equation}
L_{i}=\pi _{ij}L_{j}\hspace{1cm}\text{for\hspace{1cm}}i\preceq j
\label{PLF1}
\end{equation}%
It should be pointed out that the family of lattices may either be infinite
if the lattice spacing $a_{n}\rightarrow 0$ and (or) the total volume $%
V\rightarrow \infty $ or finite for finite volume and finite cutoff $a_{c}$.

\subsection{Strata \protect\cite{Michel1971}, \protect\cite{Rudolph1}, 
\protect\cite{VilelaStrata}}

Let $G$ be a compact Lie group acting on a manifold $\mathcal{M}$. The
action of $G$ on $\mathcal{M}$ leads to a stratification of $\mathcal{M}$
corresponding to the classes of equivalent \textit{orbits} $\left\{ g%
\mathcal{M};g\in G\right\} $. Let $S_{x}$ denote the \textit{isotropy (or
stabilizer) group} of $x\in \mathcal{M}$%
\begin{equation}
S_{x}=\left\{ \gamma \in G:\gamma x=x\right\}  \label{S1}
\end{equation}%
The \textit{stratum} $\Sigma \left( x\right) $ of $x$ is the set of points
having isotropy groups $G-$conjugated to that of $x$%
\begin{equation}
\Sigma \left( x\right) =\left\{ y\in \mathcal{M}:\exists \gamma \in
G:S_{y}=\gamma S_{x}\gamma ^{-1}\right\}  \label{S2}
\end{equation}%
If $G$ is a symmetry group for a physical system with states in $\mathcal{M}$%
, the \textit{configuration space} of the system is the quotient space $%
\mathcal{M}/G$ and a stratum is the set of points in $\mathcal{M}/G$ that
corresponds to orbits with conjugated isotropy groups. The map that, to each
orbit, assigns the conjugacy class of its isotropy group is called the 
\textit{type}. The set of strata carries a partial ordering of types, $%
\Sigma _{x}\subseteq \Sigma _{x^{\prime }}$ if there are representatives $%
S_{x}$ and $S_{x^{\prime }}$ of the isotropy groups such that $%
S_{x}\supseteq S_{x^{\prime }}$. The maximal element in the ordering of
types is the class of the center $Z(G)$ of $G$ and the minimal one is the
class of $G$ itself.

In gauge theories, one deals with the strata of the connections, the strata
being in one-to-one correspondence with the Howe subgroups of $G$, that is,
the subgroups that are centralizers of some subset in $G$. Given an holonomy
group $H_{\tau }$ associated to a connection $A$ of type $\tau $, the
stratum of $A$ is classified by the conjugacy class of the isotropy group $%
S_{\tau }$, that is, the centralizer of $H_{\tau }$, 
\begin{equation}
S_{\tau }=Z\left( H_{\tau }\right)  \label{2.8}
\end{equation}%
an important role being also played by the centralizer of the centralizer 
\begin{equation}
H_{\tau }^{\prime }=Z\left( Z\left( H_{\tau }\right) \right)  \label{2.9}
\end{equation}%
that contains $H_{\tau }$ itself. If $H_{\tau }^{\prime }$ is a proper
subgroup of $G$, the connection $A$ reduces locally to the subbundle $%
P_{\tau }=\left( \mathcal{M},H_{\tau }^{\prime }\right) $. \ Whether or not
all the strata types exist for the action of $G$ on $\mathcal{M}$ depends on
the structure of $\mathcal{M}$ itself. Global reduction depends on the
topology of $\mathcal{M}$, but it is always possible if $P=\left( \mathcal{M}%
,G\right) $ is a trivial bundle. $H_{\tau }^{\prime }$ is the structure
group of the \textit{maximal subbundle} associated to type $\tau $.
Therefore the types of strata are also in correspondence with types of
reductions of the connections to subbundles. If $S_{\tau }$ is the center of 
$G$ the connection is called \textit{irreducible}, all others are called 
\textit{reducible}. The stratum of the irreducible connections is called the 
\textit{generic stratum}. It is open and dense.

\subsection{Summary of results}

In Refs.\cite{VilelaJMP} and \cite{VilelaIJMPA}, by using measures that
satisfy a semigroup law, it was possible to check the consistency of the
measures at the one-plaquette level. Here the construction is extended for
plaquettes sharing common edges, which by induction implies the possibility
to construct pure gauge consistent measures in finite or infinite lattices.
This is the done in Section 2. The positivity of the transfer matrix and the
existence of a mass gap is also established. Then Section 3 explores the
construction of consistent measures when there are also fermion matter
fields in the lattice. Consistency of a particular measure is checked, here
however only at the one-plaquette level. Finally in Section 4 one discusses
the role of non-generic strata in lattice theories.

\section{Nonabelian pure gauge lattice theory: A review and some developments%
}

Here I will draw and extend the results already obtained in Refs. \cite%
{VilelaJMP} and \cite{VilelaIJMPA} concerning the construction of a
consistent measure, with also some new results concerning the positivity of
the measures as well as an alternative discussion of the mass gap.

\subsection{Consistent interacting measures}

\begin{figure}[htb]
\centering
\includegraphics[width=0.8\textwidth]{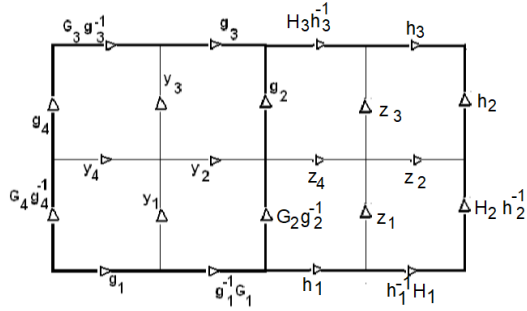}
\caption{Subdivision of two contiguous plaquettes}
\label{plaquettes_2}
\end{figure}

A state of the nonabelian theory corresponds to the assignment of an element 
$g$ of a nonabelian group $G$ to each edge of the lattice. As a reference
measure, the Haar measure of the group $G$ is also associated to each edge.
The set of independent Haar measures in the edges clearly establish a
consistent family of measures in the lattice and therefore there also exists
a reference Haar measure in the projective limit. In addition the theory is
invariant under a direct product group%
\begin{equation}
\mathcal{G}=\underset{vertices}{\Pi }G_{i}  \label{CM1}
\end{equation}%
that is, an independent copy of $G$ at each vertex which acts on the $ij$
edge group variable $g$ as%
\begin{equation}
g\rightarrow g_{i}gg_{j}^{-1}  \label{CM2}
\end{equation}%
The next step is to find nontrivial measures, that is measures that couple
the group elements in different edges, but that satisfy the consistency
requirement along the projective lattice family. In particular one looks for
densities that multiply the reference Haar measure. In \cite{VilelaJMP} and 
\cite{VilelaIJMPA} this was achieved by considering the construction of the
projective family in such a way that at each step only one plaquette is
subdivided, together with the adding of new plaquettes. However the
procedure is much more general, with the same consistency condition being
obtained for the measure densities. If the subdivided plaquettes have no
edge in common the proof of the condition for consistency of the measure
applies without any modification. The other possibility is when the
plaquettes that are subdivided have a common edge. Here I will show that the
measure consistency condition is unaltered. This is done by explicit
calculation. Consider the two subdivided plaquettes in Fig.\ref{plaquettes_2}

The measure density associated to the plaquettes is a central function of
the product of the group elements around the plaquettes and denote by $%
p\left( G_{1}G_{2}G_{3}^{-1}G_{4}^{-1}\right) $ and $p\left(
H_{1}H_{2}H_{3}^{-1}G_{2}^{-1}\right) $ the densities associated to the
large plaquettes and by $p^{\prime }$ the corresponding densities of the
small plaquettes. Furthermore assume that the densities, in addition to
being central functions, satisfy the following semigroup properties%
\begin{eqnarray}
\int p^{\prime }\left( G_{i}X\right) p^{\prime }\left( X^{-1}G_{j}\right)
d\mu _{H}\left( X\right) &=&p^{\prime \prime }\left( G_{i}G_{j}\right) 
\notag \\
\int p^{\prime \prime }\left( G_{i}X\right) p^{\prime \prime }\left(
X^{-1}G_{j}\right) d\mu _{H}\left( X\right) &=&p\left( G_{i}G_{j}\right)
\label{CM3}
\end{eqnarray}%
Then%
\begin{eqnarray}
&&\int p^{\prime }\left( g_{1}y_{1}y_{4}^{-1}g_{4}G_{4}^{-1}\right)
p^{\prime }\left( g_{1}^{-1}G_{1}G_{2}g_{2}^{-1}y_{2}^{-1}y_{1}^{-1}\right)
p^{\prime }\left( y_{4}y_{3}g_{3}G_{3}^{-1}g_{4}^{-1}\right) p^{\prime
}\left( y_{2}g_{2}g_{3}^{-1}y_{3}^{-1}\right)  \notag \\
&&\times p^{\prime }\left( h_{1}z_{1}z_{4}^{-1}g_{2}G_{2}^{-1}\right)
p^{\prime }\left( h_{1}^{-1}H_{1}H_{2}h_{2}^{-1}z_{2}^{-1}z_{1}^{-1}\right)
p^{\prime }\left( z_{4}z_{3}h_{3}H_{3}^{-1}g_{2}^{-1}\right) p^{\prime
}\left( z_{2}h_{2}h_{3}^{-1}z_{3}^{-1}\right)  \notag \\
&&\times \prod_{i=1,\cdots ,4}d\mu _{H}\left( g_{i}\right) d\mu _{H}\left(
y_{i}\right) d\mu _{H}\left( z_{i}\right) d\mu _{H}\left( G_{i}\right)
\prod_{j=1,\cdots ,3}d\mu _{H}\left( h_{j}\right) d\mu _{H}\left(
H_{j}\right)  \notag \\
&=&\int p^{\prime }\left( X_{1}X_{4}^{-1}G_{4}^{-1}\right) p^{\prime }\left(
X_{1}^{-1}G_{1}G_{2}X_{2}\right) p^{\prime }\left(
X_{3}G_{3}^{-1}X_{4}\right) p^{\prime }\left( X_{2}^{-1}X_{3}^{-1}\right) 
\notag \\
&&\times p^{\prime }\left( Y_{1}Y_{4}^{-1}G_{2}^{-1}\right) p^{\prime
}\left( H_{1}H_{2}Y_{2}Y_{1}^{-1}\right) p^{\prime }\left(
Y_{3}H_{3}^{-1}Y_{4}\right) p^{\prime }\left( Y_{2}^{-1}Y_{3}^{-1}\right) 
\notag \\
&&\times \prod_{i=1,\cdots ,4}d\mu _{H}\left( X_{i}\right) d\mu _{H}\left(
Y_{i}\right) d\mu _{H}\left( G_{i}\right) \prod_{j=1,\cdots ,3}d\mu
_{H}\left( H_{j}\right)  \notag \\
&=&\int p^{\prime \prime }\left( X_{4}^{-1}G_{4}^{-1}G_{1}G_{2}X_{2}\right)
p^{\prime \prime }\left( X_{2}^{-1}G_{3}^{-1}X_{4}\right) p^{\prime \prime
}\left( H_{1}H_{2}Y_{2}Y_{4}^{-1}G_{2}^{-1}\right) p^{\prime \prime }\left(
Y_{2}^{-1}H_{3}^{-1}Y_{4}\right)  \notag \\
&&\times d\mu _{H}\left( X_{1}\right) d\mu _{H}\left( X_{2}\right) d\mu
_{H}\left( X_{4}\right) d\mu _{H}\left( Y_{2}\right) d\mu _{H}\left(
Y_{4}\right) \prod_{i=1,\cdots ,4}d\mu _{H}\left( G_{i}\right)
\prod_{j=1,\cdots ,3}d\mu _{H}\left( H_{j}\right)  \notag \\
&=&\int p\left( G_{3}^{-1}G_{4}^{-1}G_{1}G_{2}\right) p\left(
H_{3}^{-1}G_{2}^{-1}H_{1}H_{2}\right) \prod_{i=1,\cdots ,4}d\mu _{H}\left(
G_{i}\right) \prod_{j=1,\cdots ,3}d\mu _{H}\left( H_{j}\right)  \label{CM4}
\end{eqnarray}%
The first step uses centrality, invariance of the Haar measure and the
change of variables%
\begin{eqnarray}
X_{1}
&=&g_{1}y_{1};X_{2}=g_{2}^{-1}y_{2}^{-1};X_{3}=y_{3}g_{3};X_{4}=g_{4}^{-1}y_{4}
\notag \\
Y_{1}
&=&h_{1}z_{1};Y_{2}=h_{2}^{-1}z_{2}^{-1};Y_{3}=z_{3}h_{3};Y_{4}=g_{2}^{-1}z_{4}
\label{CM5}
\end{eqnarray}%
and the second and the third the semigroup properties (\ref{CM3}). This an
explicit check of the consistency condition (\ref{PL5}).

A quite similar construction holds if the new plaquette that is subdivided
shares other edges with other already subdivided plaquettes. By induction,
with the reasoning here and in Ref.\cite{VilelaJMP} it is established that:

\begin{proposition}
Let $\mathcal{L}$ be a (finite or infinite)projective lattice family of
compact nonabelian gauge theory with the product Haar measure as reference
measure. Then a sufficient condition for the existence of a consistent
measure in $\mathcal{L}$ is that the (plaquette) densities be central
functions satisfying the semigroup conditions (\ref{CM3}).
\end{proposition}

Notice that in (\ref{CM3}) the equality sign might be simply replaced by
"proportional to", with the scaling factor being absorbed by the measure
normalization.

The choice of the semigroup defines the particular physical theory that is
implemented (or observed) in the lattice. In \cite{VilelaJMP} and \cite%
{VilelaIJMPA} it has been checked that the heat kernel associated to the
group $G$ , having the semigroup properties (\ref{HK4}) it also approximates
at small lattice spacing the formal measure associated to the Yang-Mills
Lagrangian. It might therefore be used as a rigorous definition of this as
yet undefined theory. In this case then, the measure density associated to
each plaquette is the heat kernel $K\left( U_{\square },\beta \right) $.

From the consistency condition one sees that as the plaquettes are
subdivided along the consistent family of lattices, one should replace the $%
\beta $ parameter in the heat kernel associated to each particular plaquette
in the following way:%
\begin{eqnarray}
\beta  &\rightarrow &\beta ^{\prime }=\frac{\beta }{4}\hspace{1cm}\text{for
subdivision of one }a-\text{plaquette into }4\text{ }a/2-\text{plaquettes} 
\notag \\
\beta  &\rightarrow &\beta ^{\prime \prime }=\frac{\beta }{2}\hspace{1cm}%
\text{for subdivision of one }a-\text{plaquette into }2\text{ rectangular
plaquettes}  \notag \\
&&  \label{CM5a}
\end{eqnarray}%
Hence one has $\beta \sim a^{2}$. To obtain the relation of the $\beta $
parameter to the usual coupling constant $g$ in lattice theories one should
compare the small $\beta $ limit of the heat kernel with, for example, the
Wilson action e$^{-S_{W}}$%
\begin{equation}
S_{W}=-\frac{2}{g^{2}}\sum_{\square }\mathnormal{Re}Tr\left( U_{_{\square
}}\right)   \label{CM6}
\end{equation}%
This comparison was performed in \cite{VilelaJMP} for $SU\left( 2\right) $
and $SU\left( 3\right) $, the result being that heat kernel coefficient $%
\beta $ corresponds to the square of the coupling constant%
\begin{equation}
\beta \sim g^{2}  \label{CM7}
\end{equation}

Consistency of the measure is important not only to insure a correct
matching of the description of the physical system at all length scales, but
also to establish the existence of a continuous limit when $a\rightarrow 0$.
Notice however than in the $\beta \rightarrow 0$ limit the heat kernel (the
plaquette density) ceases to be a continuous function, meaning that the
limit measure exists but is not absolutely continuous in relation to the
product Haar measure. It is however easy to give a precise meaning to this
limiting density in the framework of a gauge projective triplet (see \cite%
{VilelaJMP} Sect.III).

Here I have been assuming uniformity of the lattice spacing $a$ at each
length scale. However, sometimes it is useful, for example for the
Hamiltonian formulation, to have a different size for one of the axis, which
one may identify as time. Then one would have $\beta _{t}$ and $\beta _{s}$
corresponding respectively to the time and and space directions. When a
plaquette is subdivided only in time direction with the space direction kept
fixed, it is the second replacement in (\ref{CM5a}) that applies.

In addition to the Wilson measure, several modified lattice measures have
been proposed in the past, either to avoid lattice artifacts or to improve
the speed of convergence in numerical calculations. Most of these improved
actions do not implement measures that are consistent in the sense
considered here. An exception are the papers by Drouffe \cite{Drouffe} and
Menotti and Onofri \cite{Menotti} who also propose the use of the heat
kernel measure, although they mostly emphasize a better convergence of the
strong coupling expansion rather than its role as a consistent measure in a
projective family. The heat kernel measure has also been used by Klimek and
Kondracki in their construction of two-dimensional QCD \cite{Klimek}.
Another advantage of the heat kernel measure is the positivity of the
transfer matrix, as has already been pointed out in the past \cite{Creutz}.
Because the explicit form of the transfer matrix is important for the
Hamiltonian formulation, the transfer matrix and the proof of positivity
will be briefly sketched in the next subsection.

\subsection{Positivity of the transfer matrix}

Given a lattice Euclidean measure, a condition for this measure to
correspond to a physical theory, with an operator representation in Hilbert
space, is the positivity of the transfer matrix. The transfer matrix
propagates the system from one time to the next. In the Hilbert space
formulation time translations are generated by the Hamiltonian. Therefore
once the positivity is proved, the Hamiltonian may be obtained by taking the
logarithm of the transfer matrix and identifying the negative of the term
linear in the lattice time-spacing as the Hamiltonian.

In the $t=0$ hyperplane, the spatial group elements at each edge $%
U_{i}\left( 0,\overrightarrow{x}\right) $ are the wave function coordinates
for the Schr\"{o}dinger picture, scalar products being defined with the Haar
measure. It is also useful to restrict (or project) the Hilbert space to
gauge-invariant functions. The transfer matrix $T$ is an operator defined
from the partition function $Z$ by%
\begin{equation}
Z=\lim_{N\rightarrow \infty }Tr\left[ T^{N}\right]  \label{CM8}
\end{equation}%
$N$ being the number of lattice spacings along the time direction. Denoting
by $U_{i}\left( na,\overrightarrow{x}\right) $ and $U_{0}\left( na,%
\overrightarrow{x}\right) $ the space-like and time-like group elements at
time $na$, the partition function may be written%
\begin{equation}
Z=\prod_{n}\int \prod_{i,\overrightarrow{x}}d\mu _{H}\left( U_{i}\left( na,%
\overrightarrow{x}\right) \right) d\mu _{H}\left( U_{0}\left( na,%
\overrightarrow{x}\right) \right) \prod_{j<l}K_{jl}\left( na,\overrightarrow{%
x},\beta _{s}\right) \prod_{l^{\prime }}K_{0l^{\prime }}\left( na,%
\overrightarrow{x},\beta _{t}\right)  \label{CM9}
\end{equation}%
$K_{\mu \nu }\left( na,\overrightarrow{x},\beta _{\bullet }\right) $ being
the heat kernel associated to the $\mu \nu -$plaquette at $\overrightarrow{x}
$, where different $\beta $ coefficients are associated to time and space
directions. From (\ref{CM9}) it follows that, denoting by $U^{(n)}$ a
generic space configuration at time $na$, the matrix elements of the
transfer matrix are%
\begin{equation}
\left\langle U^{(n)}\left\vert T\right\vert U^{(n+1)}\right\rangle =\prod_{%
\overrightarrow{x}}\prod_{j<l}K_{jl}^{1/2}\left( na,\overrightarrow{x},\beta
_{s}\right) \prod_{s}K_{0s}\left( na,\overrightarrow{x},\beta _{t}\right)
\prod_{j^{\prime }<l^{\prime }}K_{j^{\prime }l^{\prime }}^{1/2}\left( \left(
n+1\right) a,\overrightarrow{x},\beta _{s}\right)  \label{CM10}
\end{equation}%
The next step is to show the positivity of this operator%
\begin{equation}
\left\langle \Psi \left( U\right) \left\vert T\right\vert \Psi \left(
U\right) \right\rangle \geq 0  \label{CM11}
\end{equation}%
$\Psi \left( U\right) $ being gauge invariant states. From (\ref{CM10}) it
is seen that $T$ is the product of three operators%
\begin{equation*}
ABA
\end{equation*}%
of which $A$ only involves elements at a fixed time and only $B$ connects
different time hyperplanes. Therefore with $\Phi \left( U\right) =A\Psi
\left( U\right) $ it suffices to prove positivity of the $B$ operator,%
\begin{equation}
B=\prod_{\overrightarrow{x}}\prod_{s}K_{0s}\left( na,\overrightarrow{x}%
,\beta _{t}\right)  \label{CM12}
\end{equation}%
which follows from the positivity of the heat kernel of compact Lie groups.
Therefore, \textit{for a lattice theory associated to a compact Lie group G,
the transfer matrix obtained from the heat kernel measure is a positive
operator.}

An alternative proof of the positivity of the transfer matrix might involve
time-reflection positivity as in \cite{Seiler}, by choosing the $t=0$
hyperplane at mid distance between two lattice space hyperplanes and a gauge
where all the edges along the time direction are set to the group identity.
Then one sees that the time-positive and time-negative parts of the $A$
operator are symmetric and that in the $B$ operator the only component that
involves time-positive and time-negative edges does so in a symmetric way.
This latter proof would however be more general, because it also applies to
any positive linear combination of traces of plaquette operators, not only
to the heat kernel.

From the logarithm of the positive transfer matrix, an Hamiltonian may be
obtained\ as the negative of the term linear in the lattice spacing. In
particular the potential term is%
\begin{equation}
V=-\frac{1}{2\beta _{t}}\sum_{j<l}\ln K_{jl}\left( \overrightarrow{x},\beta
_{s}\right)  \label{CM13}
\end{equation}%
$K_{jl}\left( \overrightarrow{x},\beta _{s}\right) $ denoting the heat
kernel associated to the $jl-$spatial plaquette at $\overrightarrow{x}$.

\section{Hamiltonian and the mass gap}

Here one considers an Hamiltonian formulation of the lattice theory, letting
the lattice size along the time direction tend to zero, $a_{t}\rightarrow 0$%
, and the one along the space directions $a_{s}$ kept fixed. Therefore also $%
\beta _{t}\rightarrow 0$ in the consistent measure. From the previous
analysis \cite{VilelaJMP} one already knows that for small $\beta $ one
obtains the same limit as for the Wilson action. Therefore one may use for
the kinetic term the same term as in Kogut-Susskind Hamiltonian \cite{Kogut}
and for the potential term the function $V$ in (\ref{CM13})%
\begin{equation}
H=c\sum \pi _{ij}^{\alpha }\pi _{ij}^{\alpha }+V  \label{CM14}
\end{equation}%
where $c$ is a positive constant related to the coupling constants or, in
the case of an Hamiltonian constructed from the consistent measure, a
function of $\beta _{t}$ and $\beta _{s}$. The operators $\pi _{ij}^{\alpha
} $ act on the group element $U_{ij}$ of the $ij$ spatial edge by%
\begin{equation}
\left[ \pi _{ij}^{\alpha },U_{ij}\right] =-\xi ^{\alpha }U_{ij}  \label{CM15}
\end{equation}%
$\xi ^{\alpha }$ being an element of the Lie algebra and $\pi _{ij}^{\alpha
} $ may be written as $i\frac{\partial }{\partial \theta _{ij}^{\alpha }}$
if $U_{ij}=e^{i\theta _{ij}^{\alpha }\xi ^{\alpha }}$

To estimate a bound on the lowest nontrivial eigenvalue $\lambda _{1}$ one
uses the bound on the Rayleigh quotient \cite{Henrot} 
\begin{equation}
\lambda _{1}=\inf_{\psi \left( U\right) \neq 0}\frac{\int \left( \left\vert
\pi \psi \left( U\right) \right\vert ^{2}+V\left\vert \psi \left( U\right)
\right\vert ^{2}\right) d\mu \left( U\right) }{\int \left\vert \psi \left(
U\right) \right\vert ^{2}d\mu \left( U\right) }  \label{CM16}
\end{equation}

Let $\psi \left( U\right) $ be normalized, $\left( \psi \left( U\right)
,\psi \left( U\right) \right) =1$. Then with $K$ an heat kernel%
\begin{equation}
\left( \psi \left( U\right) ,K\psi \left( U\right) \right) \leq \left( \psi
\left( U\right) ,\psi \left( U\right) \right) ^{1/2}\left( K\psi \left(
U\right) ,K\psi \left( U\right) \right) ^{1/2}\leq 1  \label{CM17}
\end{equation}%
the second inequality following from the contractive semigroup property of
the heat kernel. Then%
\begin{equation*}
\left( \psi \left( U\right) ,\ln K\psi \left( U\right) \right) \leq 0
\end{equation*}%
and the potential $V$ in equation (\ref{CM13}) is $\geq 0$. Therefore to
prove that $\lambda _{1}>0$ it suffices to analyze the $V=0$ case. For the
kinetic part of the Hamiltonian the lowest gauge invariant state corresponds
to one excited plaquette. In the Hamiltonian formulation one chooses a gauge
where all the edges along the time direction are set to the identity of the
group. This is not a complete gauge, yet remaining to fix the gauge in the
space slices. There one chooses a vertex $x_{0}$ of the plaquette to be
excited and uses this point to establish a maximal tree gauge on the
space-like edges. In this gauge the plaquettes nearest to the $x_{0}$ vertex
have only one link that is not set to the identity. Therefore to excite the
plaquette is the same as to excite this link. From the parametrization of
the group elements%
\begin{equation}
U=e^{i\theta ^{\alpha }\xi ^{\alpha }}  \label{CM18}
\end{equation}%
where $\theta ^{\alpha }\in \left[ 0,2\pi \right) $ or $\left[ 0,\pi \right) 
$ for a compact group, it then follows that regularity of the compact
boundary conditions implies%
\begin{equation*}
\lambda _{1}>0
\end{equation*}

In conclusion:

\begin{proposition}
At any spatial lattice spacing, the nonabelian lattice theory with heat
kernel measure has a positive mass gap.
\end{proposition}

Compactness of the group and the contractive nature of the heat kernel
semigroup are the main ingredients leading to this result.

In \cite{VilelaIJMPA} a similar conclusion was reached using
Wentzell-Freitlin estimates associated to the ground state stochastic
process. However, because they rely on some hypotheses on the construction
of the ground state, I think that the derivation above is simpler and more
satisfactory.

\section{Nonabelian lattice gauge theory with matter fields}

In addition to the nonabelian gauge fields, physical theories also contain
matter fields which conventionally are defined to live on the vertices of
the lattice. For pure gauge theories the natural gauge invariant quantity is
the plaquette product of group elements. With fermions however the basic
element is $\overline{\psi }\gamma _{\mu }\left( \nabla _{\mu }+\nabla _{\mu
}^{\ast }\right) \psi $ with $\nabla _{\mu }$ and $\nabla _{\mu }^{\ast }$
denoting the forward and backward covariant difference operators along the $%
\mu $ coordinate,%
\begin{eqnarray}
\nabla _{\mu }\psi \left( x\right) &=&\frac{1}{a}\left( U_{\mu }\left(
x\right) \psi \left( x+a\widehat{\mu }\right) -\psi \left( x\right) \right) 
\notag \\
\nabla _{\mu }^{\ast }\psi \left( x\right) &=&\frac{1}{a}\left( \psi \left(
x\right) -U_{\mu }\left( x-a\widehat{\mu }\right) ^{\dag }\psi \left( x-a%
\widehat{\mu }\right) \right)  \label{F1}
\end{eqnarray}%
Therefore a fermion gauge measure density $\nu \left( U,\overline{\psi }%
,\psi \right) $ might be a function%
\begin{equation}
\nu \left( U,\overline{\psi },\psi \right) =f\left( \left\{ \overline{\psi }%
\gamma _{\mu }\left( \nabla _{\mu }+\nabla _{\mu }^{\ast }\right) \psi
\right\} \right)  \label{F2}
\end{equation}%
$\left\{ \overline{\psi }\gamma _{\mu }\left( \nabla _{\mu }+\nabla _{\mu
}^{\ast }\right) \psi \right\} $ denoting the set of all fermion edge
strings.

\begin{figure}[htb]
\centering
\includegraphics[width=0.8\textwidth]{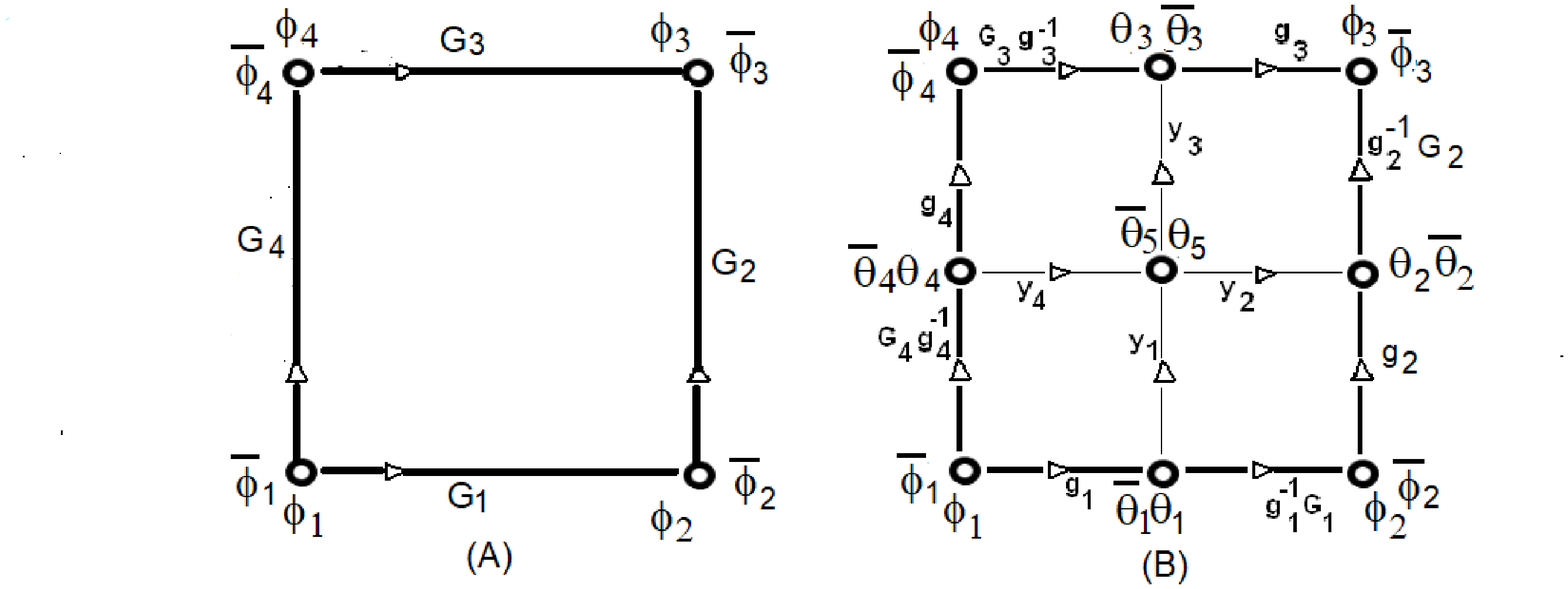}
\caption{Subdivision of a plaquette with fermion fields}
\label{plaquettes_ferm}
\end{figure}

In these strings the fermions are entities defined in a product space%
\begin{eqnarray}
\overline{\psi } &=&\overline{\chi }\otimes \overline{\phi }\in V_{\overline{%
\chi }}\otimes V_{\overline{\phi }}  \notag \\
\psi &=&\chi \otimes \phi \in V_{\chi }\otimes V_{\phi }  \label{F3}
\end{eqnarray}%
$V_{\overline{\chi }}$ and $V_{\chi }$ being Grassman spaces and $V_{%
\overline{\phi }}$ and $V_{\phi }$ representation spaces of the gauge group.
The density in (\ref{F2}) is to be multiplied by%
\begin{equation}
\prod_{i\in \text{all sites}}d\overline{\psi }_{i}d\psi _{i}\prod_{j\in 
\text{all edges}}dU_{j}  \label{F4}
\end{equation}%
Formally expanding (\ref{F2}) and by Berezin integration over the Grassman
variables$\phi $%
\begin{equation}
\prod_{j\in \text{all edges}}dU_{j}\sum_{P}(-1)^{P}\prod_{i=1\cdots N}\left.
\partial _{P(\overline{\psi }_{i}}\partial _{\psi _{i})}f\left( \left\{ 
\overline{\psi }\gamma _{\mu }\left( \nabla _{\mu }+\nabla _{\mu }^{\ast
}\right) \psi \right\} \right) \right\vert _{\overline{\psi }=\psi =0}
\label{F6}
\end{equation}%
$P$ being a permutation over all sites. The argument%
\begin{equation}
\mu =\sum_{P}(-1)^{P}\prod_{i=1\cdots N}\left. \partial _{P(\overline{\psi }%
_{i}}\partial _{\psi _{i})}f\left( \left\{ \overline{\psi }\gamma _{\mu
}\left( \nabla _{\mu }+\nabla _{\mu }^{\ast }\right) \psi \right\} \right)
\right\vert _{\overline{\psi }=\psi =0}  \label{F7}
\end{equation}%
in (\ref{F3}) is a function only of the group elements in the sites and the
edges. To obtain a consistent measure try the following ansatz%
\begin{equation}
\mu =\mu \left( \prod_{plaquettes}\overline{\phi }_{i}U_{\mu }\phi _{i+\mu }%
\overline{\phi }_{i+\mu }U_{\nu }\phi _{i+\mu +\nu }\left( \overline{\phi }%
_{i+\nu }U_{\mu }\phi _{i+\nu +\mu }\right) ^{\dag }\left( \overline{\phi }%
_{i}U_{\nu }\phi _{i+\nu }\right) ^{\dag }\right)  \label{F8}
\end{equation}%
that is, a product of plaquette-strings. Now the consistency and the choice
of the function $\mu ,$may be verified as before, here only at the single
plaquette level. Consider the four group strings around a plaquette in Fig.%
\ref{plaquettes_ferm}A. Subdividing this plaquette as in Fig.\ref%
{plaquettes_ferm}B, and denoting by $\mu ^{\prime }$ the new measure
associated to the subdivided plaquette, the functional integral becomes

\begin{eqnarray}
I &=&\int \mu ^{\prime }\left( \overline{\phi }_{1}g_{1}\theta _{1}\overline{%
\theta }_{1}y_{1}\theta _{5}\left( \overline{\theta }_{4}y_{4}\theta
_{5}\right) ^{-1}\left( \overline{\phi }_{1}G_{4}g_{4}^{-1}\theta
_{4}\right) ^{-1}\right) \mu ^{\prime }\left( \overline{\theta }%
_{1}g_{1}^{-1}G_{1}\phi _{2}\overline{\phi }_{2}g_{2}\theta _{2}\left( 
\overline{\theta }_{5}y_{2}\theta _{2}\right) ^{-1}\left( \overline{\theta }%
_{1}y_{1}\theta _{5}\right) ^{-1}\right)  \notag \\
&&\times \mu ^{\prime }\left( \overline{\theta }_{4}y_{4}\theta _{5}%
\overline{\theta }_{5}y_{3}\theta _{3}\left( \overline{\phi }%
_{4}G_{3}g_{3}^{-1}\theta _{3}\right) ^{-1}\left( \overline{\theta }%
_{4}g_{4}\phi _{4}\right) ^{-1}\right) \mu ^{\prime }\left( \overline{\theta 
}_{5}y_{2}\theta _{2}\overline{\theta }_{2}g_{2}^{-1}G_{2}\phi _{3}\left( 
\overline{\theta }_{3}g_{3}\phi _{3}\right) ^{-1}\left( \overline{\theta }%
_{5}y_{3}\theta _{3}\right) ^{-1}\right)  \notag \\
&&\times \prod_{i=1}^{5}d\mu _{H}\left( \overline{\theta }_{i}\right) d\mu
_{H}\left( \theta _{i}\right) \prod_{j=1}^{4}d\mu _{H}\left( g_{j}\right)
d\mu _{H}\left( y_{j}\right)  \label{F9}
\end{eqnarray}%
Let%
\begin{equation}
g_{1}\theta _{1}\overline{\theta }_{1}y_{1}\theta _{5}=Y_{1};\;g_{2}\theta
_{2}\overline{\theta }_{2}y_{2}^{-1}\theta _{5}=Y_{2};\;\overline{\theta }%
_{5}y_{3}\theta _{3}\overline{\theta }_{3}g_{3}=Y_{3}^{-1};\;\overline{%
\theta }_{5}y_{4}^{-1}\theta _{4}\overline{\theta }_{4}g_{4}=Y_{4}^{-1}
\label{F10}
\end{equation}%
Then using centrality, invariance of the Haar measure and integrating over
the remaining variables%
\begin{eqnarray*}
I &=&\int \mu ^{\prime }\left( \overline{\phi }_{1}Y_{1}Y_{4}^{-1}G_{4}^{-1}%
\phi _{1}\right) \mu ^{\prime }\left( Y_{1}^{-1}G_{1}\phi _{2}\overline{\phi 
}_{2}Y_{2}\right) \mu ^{\prime }\left( Y_{4}Y_{3}^{-1}G_{3}^{-1}\phi _{4}%
\overline{\phi }_{4}\right) \mu ^{\prime }\left( Y_{2}^{-1}G_{2}\phi _{3}%
\overline{\phi }_{3}Y_{3}\right) \\
&&\times \prod_{i}^{4}d\mu _{H}\left( Y_{i}\right)
\end{eqnarray*}%
If there is a family of measures $\mu ^{\prime }$, $\mu ^{\prime \prime }$
and $\mu ^{\prime \prime \prime }$ satisfying the semigroup property as
discussed before (Eq.(\ref{CM3}))%
\begin{eqnarray}
I &=&\int \mu ^{\prime \prime }\left( Y_{4}^{-1}G_{4}^{-1}\phi _{1}\overline{%
\phi }_{1}G_{1}\phi _{2}\overline{\phi }_{2}Y_{2}\right) \mu ^{\prime \prime
}\left( G_{3}^{-1}\phi _{4}\overline{\phi }_{4}Y_{4}Y_{2}^{-1}G_{2}\phi _{3}%
\overline{\phi }_{3}\right) d\mu _{H}\left( Y_{2}\right) d\mu _{H}\left(
Y_{4}\right)  \notag \\
&=&\int \mu ^{\prime \prime }\left( G_{4}^{-1}\phi _{1}\overline{\phi }%
_{1}G_{1}\phi _{2}\overline{\phi }_{2}Y_{2}Y_{4}^{-1}\right) \mu ^{\prime
\prime }\left( Y_{4}Y_{2}^{-1}G_{2}\phi _{3}\overline{\phi }%
_{3}G_{3}^{-1}\phi _{4}\overline{\phi }_{4}\right) d\mu _{H}\left(
Y_{2}\right) d\mu _{H}\left( Y_{4}\right)  \notag \\
&=&\mu \left( G_{4}^{-1}\phi _{1}\overline{\phi }_{1}G_{1}\phi _{2}\overline{%
\phi }_{2}G_{2}\phi _{3}\overline{\phi }_{3}G_{3}^{-1}\phi _{4}\overline{%
\phi }_{4}\right)  \label{F11}
\end{eqnarray}%
Hence the consistency condition on the measures may also be fulfilled with
fermions on the lattice, by using central measures with the semigroup
property. The consistency condition was verified after integration over the
Grassman variables. The full measure in (\ref{F7}) is implicitly defined as
the measure that by integration over the Grassman part of the fermion
variables leads to (\ref{F8}).

So far, in the derivation leading to the consistency condition in Eq.(\ref%
{F11}), $G_{i},g_{i},y_{i},\phi _{i},\overline{\phi }_{i},\theta _{i},%
\overline{\theta }_{i}$ have been considered arbitrary group elements.
However specification of a physical theory requires the choice of particular
representations for these group elements. In $SU\left( n\right) $ non
Abelian gauge theories is usual to choose the defining $n^{2}-1$ dimensional
representation for the group elements associated to the edges of the
lattice. Then to the group elements $\phi _{i},\overline{\phi }_{i},\theta
_{i},\overline{\theta }_{i}$ at the vertices one may associate the
fundamental $n$ and $\overline{n}$ representations. These elements always
appear in the measure in the $\phi _{i}\overline{\phi }_{i}$ and $\theta _{i}%
\overline{\theta }_{i}$ combinations which will decompose into a scalar and
a $n^{2}-1$ dimensional representation. Hence the same measure may contain
both the pure gauge part and the matter fields.

\section{Strata and the lattice potential}

The measures, that have been discussed before, provide the probability of
each particular group configuration in the lattice. In particular they
provide the integration measure that controls the fluctuations around the
ground state. Let us parametrize the group elements in the usual way%
\begin{equation}
U_{\mu }\left( x\right) =e^{\int_{x}^{x+\widehat{\mu }}ds\tau _{a}A_{a}^{\mu
}\left( s\right) }  \label{ST1}
\end{equation}%
Of particular interest are ground state configurations corresponding to
condensates. Invariance of the measure implies that the fluctuations are
around zero mean,%
\begin{equation}
\left\langle A_{a}^{\mu }\right\rangle =\left\langle F_{a}^{\mu \nu
}\right\rangle =0  \label{ST2}
\end{equation}%
$F_{a}^{\mu \nu }$ being the plaquette field, while quantities like $%
\left\langle A_{a}^{\mu }A_{a\mu }\right\rangle $ and $\left\langle
F_{a}^{\mu \nu }F_{a\mu \nu }\right\rangle $ may be different from zero.

The multiplicity of backgrounds may be related to the strata of the gauge
group operating in the lattice. The configuration space of a pure gauge
lattice $L_{i}$ (associated to the group $G$) in the projective family $%
\mathcal{L}$ is%
\begin{equation}
\mathcal{M=}G^{\otimes N}  \label{ST3}
\end{equation}%
$N$ being the number of edges in the lattice. On $\mathcal{M}$ acts the
gauge group%
\begin{equation}
\mathcal{G=}G^{\otimes N_{0}}  \label{ST4}
\end{equation}%
$N_{0}$ being the number of vertices in the lattice. The action of the group
may be reduced from $G^{\otimes N_{0}}$ to $G$ by choosing a maximal tree
gauge \cite{Rudolph2}, connecting a particular vertex $x_{0}$ to all
vertices in the lattice and assigning to the identity all group elements in
the edges along the tree. The remaining edges not in the \ tree have both
vertices group-identified (more precisely, parallel transported) to the
point $x_{0}$. Therefore the reduced system becomes%
\begin{equation}
\mathcal{M}^{\prime }\mathcal{=}G^{\otimes M}  \label{ST5}
\end{equation}%
where in general $M<<N$ and in $\mathcal{M}^{\prime }$ acts the $G$ group by
the conjugate action%
\begin{equation}
\left( g_{1},g_{2},\cdots ,g_{M}\right) \rightarrow \left(
gg_{1}g^{-1},gg_{2}g^{-1},\cdots ,gg_{M}g^{-1}\right)  \label{ST6}
\end{equation}%
$g\in G$. $\mathcal{M}^{\prime }$ is effectively a set of $M$ $\ x_{0}-$
based loops acted upon by $G$

The strata of the lattice configuration space are the strata of the action
of $G$ on $G^{M}$. They are at most as many as the number of Howe subgroups
of $G$ and for $G=SU\left( n\right) $ they were fully characterized by the
authors of refs. \cite{Rudolph2} \cite{Rudolph5}.

Let, for example, $G=SU\left( 3\right) $. Let each orbit be characterized by
the pair $\left( \lambda _{1},\lambda _{2}\right) $ of eigenvalues of the
group elements (the two independent diagonal elements of the maximal torus)
and $M\geq 2$. If the set $\left( g_{1},g_{2},\cdots ,g_{M}\right) $ has no
common eigenspace the stabilizer subgroup is the center of $SU\left(
3\right) $ and this is the generic stratum. If there is one common
eigenspace the stabilizer is $U\left( 1\right) $. If there are three
different common one-dimensional eigenspaces the stabilizer is $U\left(
1\right) \times U\left( 1\right) $ and if there is one common
two-dimensional eigenspace the stabilizer is $U\left( 2\right) $. Finally if
there is a common three-dimensional eigenspace, meaning that all elements in 
$\left( g_{1},g_{2},\cdots ,g_{M}\right) $ are the identity the stabilizer
is $SU\left( 3\right) $. Hence there are $5$ different strata.

Classical dynamics in the lattice takes place in the phase space, the
cotangent bundle $T\mathcal{M}^{\prime \ast }=TG^{\otimes M\ast }$. If the
initial condition lies in an orbit of a particular stratum, the classical
system remains there for all its undisturbed classical evolution. Therefore
for classical dynamics it makes sense to consider and classify dynamics in
the different strata although, for random initial conditions, the generic
strata will be almost surely chosen.

However for quantum mechanics, the situation is different because the wave
function will surely explore different strata and, the generic stratum
having full measure, it would seem that it is only the generic stratum that
matters. Nevertheless, some authors \cite{Cobra} \cite{Emmrich} have argued
that in systems with gauge symmetry, where the configuration space is a
orbifold with singularities corresponding to points of non-generic higher
symmetry, one may find concentrations of the wave functions near the
non-generic strata. This may depend on the form of the Hamiltonian used in
the Schr\"{o}dinger equation.

The role of the non-generic strata in the Hamiltonian formulation of the
lattice theory may be analyzed with the Hamiltonian (\ref{CM14}) and the
potential (\ref{CM13}). Let for definiteness $G=SU\left( 3\right) $. The
heat kernel that enters the potential (\ref{CM13}) is a function of the two
angles $\left( \theta _{1},\theta _{2}\right) $ in the maximal torus $%
diag\left\{ \exp \left( i\theta _{1}\right) ,\exp \left( i\theta _{2}\right)
,\exp \left( -i\left( \theta _{1}+\theta _{2}\right) \right) \right\} $ of
the group element associated to each plaquette. When the lattice is refined
to small lattice spacing the heat kernel becomes \cite{VilelaJMP} 
\begin{equation}
K\left( \beta _{s}\right) \underset{\beta _{s}\rightarrow 0}{\longrightarrow 
}\exp \left\{ -\frac{1}{2\beta _{s}}\left( \theta _{1}^{2}+\theta
_{2}^{2}+\theta _{1}\theta _{2}\right) \right\}  \label{ST7}
\end{equation}%
that is, it contributes to the potential an harmonic term $\frac{1}{4\beta
_{t}\beta _{s}}\left( \theta _{1}^{2}+\theta _{2}^{2}+\theta _{1}\theta
_{2}\right) $.

Consider now a spatial lattice of dimension $N\times N\times N$. A maximal
tree gauge starting for an upper corner is essentially equivalent to an
axial gauge $A_{3}=0$ $\left( U_{3}=1\right) $. There are then $2\times
N^{3} $ independent edge group elements, and $3\times N^{3}$ plaquette group
elements, not all independent. Of these there are $2\times N^{3}$ plaquettes
along the $3$ direction with two non-trivial edges each and $N^{3}$
plaquettes along the $1,2$ planes with four nontrivial edges.

After the gauge fixing each edge group element is still acted upon by the
conjugate action as in (\ref{ST6}) and the classification of the strata is
as described before. Consider now the non-generic stratum with stabilizer $%
U\left( 1\right) $. With a common eigenvalue in all independent links one
obtains $\theta _{1}=0$ for all the plaquettes. Therefore for each fixed $%
\left\{ \theta _{2}^{(ij)}\right\} $ configuration one obtains a minimum of
the potential on this stratum. Likewise, in the strata with $SU\left(
2\right) $ stabilizer one obtains $\theta _{1}=\theta _{2}=0$ for the
plaquettes, another potential minimum. Hence, in the potential of this high
dimensional Schr\"{o}dinger problem there are multiple different local
minima associated to the non-generic strata. It is intriguing to realize
that this multiple minima situation is the one that might lead to a fast
growing point spectrum for the gauge backgrounds, as shown in \cite{Fast}
using an inverse scattering argument.

\end{document}